\documentclass[sigconf]{acmart}

\AtBeginDocument{%
  \providecommand\BibTeX{{%
    \normalfont B\kern-0.5em{\scshape i\kern-0.25em b}\kern-0.8em\TeX}}}

\setcopyright{rightsretained} 
\copyrightyear{2023} 
\acmYear{2023} 
\acmDOI{10.1145/3544548.3580984} 

\acmConference[CHI '23]{Proceedings of the 2023 CHI Conference on Human Factors in Computing Systems}{April 23--28, 2023}{Hamburg, Germany}
\acmBooktitle{Proceedings of the 2023 CHI Conference on Human Factors in Computing Systems (CHI '23), April 23--28, 2023, Hamburg, Germany}
\acmISBN{978-1-4503-9421-5/23/04}

\usepackage[inline,shortlabels]{enumitem} 
\usepackage{tabularx}

\begin{document}

\title[Contestable Camera Cars]{Contestable Camera Cars: A Speculative Design Exploration of Public AI That Is Open and Responsive to Dispute}

\author{Kars Alfrink}
\email{c.p.alfrink@tudelft.nl}
\orcid{0000-0001-7562-019X}
\affiliation{%
  \institution{Delft University of Technology}
  \department{Department of Sustainable Design Engineering}
  \streetaddress{Landbergstraat 15}
  \city{Delft}
  \country{The Netherlands}
  \postcode{2628 CE}}

\author{Ianus Keller}
\email{a.i.keller@tudelft.nl}
\affiliation{%
  \institution{Delft University of Technology}
  \department{Department of Human-Centered Design}
  \streetaddress{Landbergstraat 15}
  \city{Delft}
  \country{The Netherlands}
  \postcode{2628 CE}}

\author{Neelke Doorn}
\email{n.doorn@tudelft.nl}
\affiliation{%
  \institution{Delft University of Technology}
  \department{Department of Values, Technology and Innovation}
  \streetaddress{Jaffalaan 5}
  \city{Delft}
  \country{The Netherlands}
  \postcode{2628 BX}}

\author{Gerd Kortuem}
\email{g.w.kortuem@tudelft.nl}
\affiliation{%
  \institution{Delft University of Technology}
  \department{Department of Sustainable Design Engineering}
  \streetaddress{Landbergstraat 15}
  \city{Delft}
  \country{The Netherlands}
  \postcode{2628 CE}}

\renewcommand{\shortauthors}{Alfrink, Keller, Doorn, and Kortuem}

\begin{abstract}
    Local governments increasingly use artificial intelligence (AI) for automated decision-making. 
    Contestability, making systems responsive to dispute, is a way to ensure they respect human rights to autonomy and dignity. 
    We investigate the design of public urban AI systems for contestability through the example of camera cars: human-driven vehicles equipped with image sensors. 
    Applying a provisional framework for contestable AI, we use speculative design to create a concept video of a contestable camera car. 
    Using this concept video, we then conduct semi-structured interviews with 17 civil servants who work with AI employed by a large northwestern European city. 
    The resulting data is analyzed using reflexive thematic analysis to identify the main challenges facing the implementation of contestability in public AI. 
    We describe how civic participation faces issues of representation, public AI systems should integrate with existing democratic practices, and cities must expand capacities for responsible AI development and operation.
\end{abstract}




\maketitle

\section{Introduction}

Local governments increasingly use artificial intelligence (AI) to support or entirely automate public service decision-making. We define AI broadly, following \citet{suchman_corporate_2018}: ``[a] cover term for a range of techniques for data analysis and processing, the relevant parameters of which can be adjusted according to either internally or externally generated feedback.'' As the use of AI in public sector decision-making increases, so do concerns over its harmful social consequences, including the undermining of the democratic rule of law and the infringement of fundamental human rights to dignity and self-determination \citep[e.g.][]{chiusi_automating_2020, crawford_ai_2019}. Increasing systems' \emph{contestability} is a way to counteract such harms. Contestable AI is a small but growing field of research \cite{hirsch_designing_2017, almada_human_2019, vaccaro_contestability_2019, sarra_put_2020, henin_beyond_2021, alfrink_contestable_2022}.
However, the contestable AI literature lacks guidance for application in specific design situations. In general, designers need \emph{examples} and \emph{instructions} to apply a framework effectively \cite{hook_strong_2012, lowgren_annotated_2013}. We, therefore, seek to answer the questions: RQ1: What are the characteristics of a contestable public AI system? RQ2: What are the challenges facing the implementation of contestability in public AI?

We ground our work in the use of camera cars: human-driven vehicles equipped with image sensors used for \textit{vehicular urban sensing} (VUS). The primary motivation for these systems is increased efficiency (cost reduction), for example for parking enforcement. Outside of the densest urban areas, costs of traditional means of parking enforcement quickly exceed collected fees \cite{mingardo_rotterdam_2020}.
Ethical concerns over using camera cars for these and other purposes reflect those around smart urbanism more broadly: data is captured without consent or notice, its benefits favor those doing the capturing, leading to reductionist views and overly technocratic decision-making \cite{kitchin_ethics_2016}.

In this paper, we explore the shape contestable AI may take in the context of local government public services and we describe the responses of civil servants who work with AI to these future visions.

Our design methods are drawn from speculative, critical and future-oriented approaches \cite{dunne_speculative_2013, bardzell_reading_2014, knutz_role_2014, galloway_speculative_2018}. We use the `Contestable AI by Design' framework \cite{alfrink_contestable_2022} as a generative tool to design a concept for a contestable camera car system. Using the resulting concept video as a prompt, we conduct semi-structured interviews with civil servants employed by Amsterdam who work with AI. Our focus here is on the challenges our respondents see towards implementing these future visions and contestability more generally. We then use reflexive thematic analysis \cite{braun_using_2006, cooper_thematic_2012, braun_thematic_2021} to generate themes from the interview transcripts that together describe the major challenges facing the implementation of contestability in public AI.\footnote{This study was preregistered at Open Science Framework: \url{https://osf.io/26rts}}

The empirical work for this study was conducted in Amsterdam. The city has previously explored ways of making camera cars more ``human-friendly.'' But efforts so far have been limited to up-front design adjustments to camera cars' physical form.\footnote{\url{https://responsiblesensinglab.org/projects/scan-cars}}

The contributions of this paper are twofold: First, we create an example near-future concept of a contestable AI system in the context of public AI, specifically camera-based VUS. The concept video is usable for debating the merits of the contestable AI concept and exploring implications for its implementation. Second, we offer an account of the challenges of implementing contestability in public AI, as perceived by civil servants employed by Amsterdam who work with AI.

We structure this paper as follows: First, we introduce Amsterdam and its current use of camera cars for parking enforcement and other purposes. Next, we discuss related work on contestable AI, public and urban AI, VUS, and speculative design. Subsequently, we describe our research approach, including our design process, interview method, and data analysis. We then report on the resulting design concept and civil servant responses. Finally, we reflect on what our findings mean for current notions of contestable AI and consider the implications for its design in the context of public and urban AI in general and camera-based VUS in particular.

\section{Background}

\subsection{Amsterdam}\label{sec:amsterdam}

Amsterdam is the capital and largest city of the Netherlands. Its population is around 0.9 million (881.933 in 2022).\footnote{\url{https://onderzoek.amsterdam.nl/interactief/kerncijfers}} ``By Dutch standards, the city is a financial and cultural powerhouse'' \cite{raven_urban_2019}.

Amsterdam is intensely urbanized. The city covers 219.492 km2 of land (2019). The city proper has 5.333 (2021) inhabitants per km2 and 2.707 (2019) houses per km2.\footnote{\url{https://onderzoek.amsterdam.nl/interactief/kerncijfers}} Amsterdam is considered the financial and business capital of the country. It is home to a significant number of banks and corporations. Its port is the fourth largest in terms of sea cargo in Northwest Europe.\footnote{\url{https://www.amsterdam.nl/bestuur-organisatie/volg-beleid/economie/haven}} Amsterdam is also one of the most popular tourist destinations in Europe.\footnote{\url{https://onderzoek.amsterdam.nl/publicatie/bezoekersprognose-2022-2024}}

In 2022, over a third (35\%) of residents were born abroad.\footnote{\url{https://onderzoek.amsterdam.nl/interactief/dashboard-kerncijfers}}
Amsterdam has relatively many households with a very low income (17\%) and a very high income (14\%).\footnote{\url{https://onderzoek.amsterdam.nl/publicatie/de-staat-van-de-stad-amsterdam-xi-2020-2021}}
In 2020, Amsterdam's working population (age 15-74) was relatively highly educated (48\%).\footnote{\url{https://onderzoek.amsterdam.nl/publicatie/de-staat-van-de-stad-amsterdam-xi-2020-2021}}

The city is governed by a directly elected municipal council, a municipal executive board, and a government-appointed mayor. The mayor is a member of the board but also has individual responsibilities.
The 2022-2026 coalition agreement's final chapter on ``cooperation and organization'' contains a section on ``the digital city and ICT,'' which frames technology as a way to improve services and increase equality and emancipation. Among other things, this section focuses on protecting citizens' privacy, safeguarding digital rights, monitoring systems using an algorithm register\footnote{\url{https://algoritmeregister.amsterdam.nl}}, testing systems for ``integrity, discrimination and prejudice'' throughout their lifecycle, and the continuing adherence to principles outlined in a local manifesto describing values for a responsible digital city\footnote{\url{https://tada.city}}.

\subsection{Camera car use in Amsterdam}

In January 2021, 13 municipalities in the Netherlands, including Amsterdam, made use of camera cars for parking monitoring and enforcement.\footnote{\url{https://www.rtlnieuws.nl/nieuws/nederland/artikel/5207606/scanauto-boete-aanvechten-grote-steden-amsterdam-utrecht-den-haag}}

Paid parking targets parking behavior and car use of citizens, businesses, and visitors. Its aims are to reduce the number of cars in the city, relieve public space pressures, and improve air quality. Cities expect to make alternative modes of transportation (cycling, public transport) more attractive by charging parking fees and limiting the availability of parking licenses per area.

The system in Amsterdam checks if parked cars have paid their parking fee or have a parking permit. Community service officers use cars outfitted with cameras to patrol city parking areas. They capture images of license plates and use computer vision algorithms to recognize license plates. The system uses these license plates to check with a national parking register if a vehicle has the right to park in its location and at the given time. Payment must be made within 5 minutes after the vehicle has been `scanned.' If not, a parking inspector employed by the company that operates the system on behalf of the city reviews the situation based on four photos to determine if exceptional circumstances apply (e.g., curbside (un)loading, stationary at traffic light). This human reviewer also checks if the license plate is recognized correctly. In case of doubt, they dispatch a parking controller by motor scooter to assess the situation on-site. The system issues a parking fine if no exceptional circumstances apply by passing the data to the municipal tax authorities. They then use the same parking register database to retrieve the personal data of the owner of the vehicle to send them a parking fine.

A dedicated website allows people to appeal a fine within six weeks of issuing. The website provides access to the environment and license plate photos. (Any bystanders, unrelated license plates, and other privacy-sensitive information are made unrecognizable.) A third-party service also offers to object to traffic and parking fines on behalf of people, free of charge.

Amsterdam also uses the parking monitoring camera cars to detect stolen vehicles and vehicles with a claim from the police or the public prosecutor. Cars are registered as stolen in the parking register. In case of a match with a scanned license plate, a national vehicle crime unit, possibly cooperating with the police, can take action. Data is also collected about `parking pressure' and the types of license holders for municipal policy development.

Finally, Amsterdam is exploring additional applications of camera cars, including outdoor advertisement taxes\footnote{\url{https://responsiblesensinglab.org/projects/scan-cars}} and side-placed garbage collection.\footnote{\url{https://medium.com/maarten-sukel/garbage-object-detection-using-pytorch-and-yolov3-d6c4e0424a10}}

\section{Related work}

\subsection{Contestable AI by design}

A small but growing body of research explores the concept of contestable AI \cite{hirsch_designing_2017, almada_human_2019, vaccaro_contestability_2019, sarra_put_2020, henin_beyond_2021, alfrink_contestable_2022}. Contestability helps to protect against fallible, unaccountable, unlawful, and unfair automated decision-making. It does so by ensuring the possibility of human intervention throughout the system lifecycle, and by creating arenas for adversarial debate between decision subjects and system operators.

\citet{hirsch_designing_2017} define contestability as ``humans challenging machine predictions,'' framing it as a way to protect against inevitably fallible machine models, by allowing human controllers to intervene before machine decisions are put into force.
\citet{vaccaro_contestability_2019} frame contestability as a ``deep system property,'' representing joint human-machine decision-making. Contestability is a form of procedural justice, giving voice to decision subjects, and increasing perceptions of fairness.
\citet{almada_human_2019} defines contestability as the possibility for ``human intervention,'' which can occur not only post-hoc, in response to an individual decision, but also ex-ante, as part of AI system development processes. For this reason, they argue for a practice of ``contestability by design.''
\citet{sarra_put_2020} argues that contestability exceeds mere human intervention. They argue that contestability requires a ``procedural relationship.'' A ``human in the loop'' is insufficient if there is no possibility of a ``dialectical exchange'' between decision subjects and human controllers.
Finally, \citet{henin_beyond_2021} argue that the absence of contestability undermines systems' legitimacy. They distinguish between explanations and justifications. The former are descriptive and intrinsic to systems themselves. The latter are normative and extrinsic, depending on outside references for assessing outcomes' desirability. Because contestability seeks to show that a decision is inappropriate or inadequate, it requires justifications, in addition to explanations.

Building on these and other works, \citet{alfrink_contestable_2022} define contestable AI as ``open and responsive to human intervention, throughout their lifecycle, establishing a procedural relationship between decision subjects and human controllers.'' They develop a preliminary design framework that synthesizes elements contributing to contestability identified through a systematic literature review. The framework comprises five system features and six development practices, mapped to major system stakeholders and typical AI system lifecycle phases. For \citet{alfrink_contestable_2022}, contestability is about ``leveraging conflict for continuous system improvement.''

Most of the works \citet{alfrink_contestable_2022} include are theoretical rather than empirical, and are not derived from specific application contexts. Contexts that do feature in works discussed are healthcare \cite{hirsch_designing_2017, ploug_four_2020}, smart cities \cite{jewell_contesting_2018}, and content moderation \cite{elkin-koren_contesting_2020, vaccaro_at_2020, vaccaro_contestability_2021}. The framework has not been validated, and lacks guidance and examples for ready application by practitioners.

\subsection{Public \& urban AI}

An increasing number of researchers report on studies into the use of AI in the public sector, i.e., \textit{public AI} \cite{brown_toward_2019, dalpiaz_trustworthy_2021, de_fine_licht_artificial_2020, fatima_public_2022, flugge_perspectives_2021, marda_data_2020, pine_politics_2015, saxena_conducting_2020, saxena_framework_2021, veale_fairness_2018}.
Although some do use the term ``AI'' \cite{dalpiaz_trustworthy_2021, de_fine_licht_artificial_2020, fatima_public_2022}, more commonly the term used is ``algorithm'' or ``algorithmic system'' \cite{brown_toward_2019, saxena_framework_2021, flugge_perspectives_2021, saxena_conducting_2020, veale_fairness_2018, pine_politics_2015}.
These algorithmic systems are put to use for informing or automating (public) decision-making by government public service (or sector) agencies \cite{brown_toward_2019, dalpiaz_trustworthy_2021, fatima_public_2022, saxena_conducting_2020, saxena_framework_2021}.
The application contexts researchers report on include: child protection \cite{brown_toward_2019, saxena_conducting_2020, saxena_framework_2021, veale_fairness_2018, dalpiaz_trustworthy_2021}; public housing \cite{dalpiaz_trustworthy_2021}; public health \cite{dalpiaz_trustworthy_2021, pine_politics_2015}; social protection \cite{flugge_perspectives_2021, dalpiaz_trustworthy_2021, veale_fairness_2018}; public security \cite{marda_data_2020, fatima_public_2022} and taxation \cite{veale_fairness_2018}.
Some of the issues explored include: how transparency, explanations and justifications may affect citizens' trust, acceptance and perceived legitimacy of public AI \cite{brown_toward_2019, dalpiaz_trustworthy_2021, de_fine_licht_artificial_2020}; the politics of measurement, the human subjective choices that go into data collection, what does and does not get counted, and in what way \cite{marda_data_2020, pine_politics_2015}; and how public sector employees' work is impacted by public AI \cite{flugge_perspectives_2021}, with a particular focus on discretion \cite{saxena_conducting_2020, saxena_framework_2021}, and how research and practice might more productively collaborate \cite{veale_fairness_2018}.

An overlapping but distinct area of research focuses on the role of AI in the built environment, so-called \textit{urban AI} \cite{alfrink_tensions_2022, howe_integrative_2022, luusua_artificial_2020, luusua_urban_2020, sawhney_contestations_2022, tseng_assemblage_2022, yigitcanlar_artificial_2022}. Many application contexts here are mobility-related, for example smart electric vehicle charging \cite{alfrink_tensions_2022}; autonomous vehicles \cite{luusua_artificial_2020}; and automated parking control systems \cite{sawhney_contestations_2022}. The focus of this research tends to be more on how AI molds, mediates, and orchestrates the daily lived experience of urban places and spaces. Ethical questions related to AI become intertwined with city-making ethics, ``who has the right to design and live in human environments'' \cite{luusua_urban_2020}. What the urban AI `lens' adds to public AI discourse are questions of \textit{spatial justice} \cite{soja_city_2009} in addition to those of procedural and distributive justice.

\subsection{Vehicular urban sensing}

Vehicular (urban) sensing is when ``vehicles on the road continuously gather, process, and share location-relevant sensor data'' \cite{lee_survey_2010}. They are ``a prominent example of cyber-physical systems'' requiring a multidisciplinary approach to their design \cite{pau_challenges_2012}. Sensors can be mounted on vehicle, or onboard smartphones may be used instead or in addition \cite{fan_towards_2021, lee_survey_2010}. Vehicles, here, are usually cars (automobiles). One advantage of cars is that they have few power constraints \cite{pau_challenges_2012}. Much of the literature to date focuses on enlisting privately owned vehicles in crowdsourcing efforts \cite{fan_towards_2021, lai_optimized_2022, pau_challenges_2012, zuo_cost-effective_2017}, as well as networking infrastructure challenges \cite{lee_survey_2010, pau_challenges_2012, zuo_cost-effective_2017, bruno_efficient_2015}. A wide range of sensors is discussed, but some focus specifically on the use of cameras \cite{bloodworth_using_2015, bruno_efficient_2015, mingardo_rotterdam_2020, zhang_hybrid_2007}. Applications include traffic monitoring and urban surveillance \cite{bruno_efficient_2015}, air pollution and urban traffic \cite{pau_challenges_2012}, infrastructure monitoring (i.e., ``remote assessment of structural performance'') \cite{bloodworth_using_2015}, and (of particular note for our purposes here) parking monitoring and enforcement \cite{mingardo_rotterdam_2020}. \citet{mingardo_rotterdam_2020} describes enforcement of on-street parking in Rotterdam, the Netherlands, using ``scan-cars.'' They claim the main reason for introducing this system was to reduce the cost of enforcement. Income usually covers enforcement costs in areas with high fees and large numbers of motorists. However, residents usually have affordable parking permits in peripheral areas, and the area to cover is much larger. Systems like the one in Rotterdam use so-called ``automatic number plate recognition'' (ANPR). \citet{zhang_hybrid_2007} propose an approach to segmenting license plates that can deal with a wide range of angles, lighting conditions, and distances. They report an accuracy of 95\%.

\subsection{Speculative design}

We use `speculative design' as a cover term for various forms of design futuring, including design fiction and critical design.
Speculative design seeks to represent or ``project'' future consequences of a current issue \cite{disalvo_design_2009}.

Although early exemplars of speculative design often took the form of products, later projects usually include various forms of storytelling, primarily to aid audience interpretation and engagement \cite{galloway_speculative_2018}. \citet{auger_speculative_2013} calls this a design's ``perceptual bridge.''
\citet{sterling_cover_2009} frames design fiction as a marriage of science-fiction literature and industrial product design, which should address the inabilities of both to ``imagine effectively.''
\citet{kirby_future_2010} has described the relationship between science-fiction cinema and design. Design in service of cinema produces ``diegetic prototypes,'' objects that function within a film's story world.
Alternatively, as \citet{bleecker_design_2009} puts it, speculative design produces things that tell stories and, in the audience's minds, create future worlds.
This notion is similar to what \citet{dunne_speculative_2013} call ``design as a catalyst for social dreaming.''
For them, the focus of speculation is on the implications of new developments in science and technology.
As such, they claim speculative design can contribute to new ``sociotechnical imaginaries'' \cite{jasanoff_containing_2009, jasanoff_dreamscapes_2015}.

Speculative design can be a way to ``construct publics'' around ``matters of concern'' \cite{disalvo_design_2009, bleecker_design_2009, forlano_design_2014}, to ``design for debate'' \cite{malpass_criticism_2015}. It is about asking questions rather than solving problems \cite{forlano_design_2014, galloway_speculative_2018}. 
To spark debate, speculative design must be provocative \cite{bardzell_critical_2012}.
It evokes critical reflection using satirical wit \cite{malpass_between_2013}.
For this satire to work, the audience must read speculative designs \emph{as} objects of design, contextualized and rationalized with a narrative of use \cite{malpass_between_2013, malpass_criticism_2015}.
Speculative designs do not lack function and can, therefore, not be dismissed as mere art. Instead, speculative design leverages a broader conception of function that goes beyond traditional notions of utility, efficiency, and optimization and instead seeks to be relational and dynamic \cite{malpass_criticism_2015}. 

To further support audiences' engagement in debate, some attempts have combined speculative design with participatory approaches. 
In workshop-like settings, speculative designs co-created with audiences can surface controversies, and be a form of ``infrastructuring'' that creates ``agonistic spaces'' \cite{hillgren_prototyping_2011, forlano_design_2014, galloway_speculative_2018}.

Early work was primarily focused on speculative design as a `genre,' exploring what designs can do, and less on how it should be practiced \cite{galloway_speculative_2018}. Since then, some have explored speculative design as a method in HCI design research, particularly in `research through design' or `constructive design research' \cite{bardzell_critical_2012, bardzell_what_2013, galloway_speculative_2018}.

There have been a few attempts at articulating criteria by which to evaluate speculative designs \cite{disalvo_design_2009, bardzell_what_2013, bardzell_reading_2014, galloway_speculative_2018}. 
Some works offer guidelines for what makes speculative design critical \cite{bardzell_what_2013}; reflecting on speculative designs \cite{kozubaev_expanding_2020}; evaluations that match expected knowledge outcomes \cite{baumer_evaluating_2020}; and `tactics' for that drawn from a canon of exemplars \cite{ferri_analyzing_2014}.

\section{Method}

Our overall approach can be characterized as constructive design research that sits somewhere between what \citet{koskinen_design_2011} calls the `field' and `showroom' modes or research through design using the `genre' of speculative design \cite{galloway_speculative_2018}. We create a concept video of a near future contestable camera car. We actively approach our audience to engage with the concept video through interviews. We use storytelling to aid audience interpretation, to help them recognize how a contestable camera car might fit in daily life. We seek to strike a balance between strangeness and normality. We measure success by the degree to which our audience is willing and able to thoughtfully engage with the concept video. In other words, we use speculative design to ask questions, rather than provide answers.

Our study is structured as follows: (1) we first formulate a design brief to capture the criteria that the speculative design concept video must adhere to; (2) we then conduct the speculative design project; (3) a rough cut of the resulting concept video is assessed with experts; (4) the video is then adjusted and finalized; (5) using the final cut of the speculative design concept video as a `prompt' we then conduct semi-structured interviews with civil servants; (6) finally, we use the interview transcripts for reflexive thematic analysis, exploring civil servants' views of challenges facing the implementation of contestability.

The data we generate consist of: (1) visual documentation of the design concepts we create; and (2) transcripts of semi-structured interviews with respondents. The visual documentation is created by the principal researcher and design collaborators as the product of the design stage. The transcripts are generated by an external transcriber on the basis of audio recordings.

\subsection{Design process}

We first created a design brief detailing assessment criteria for the design outcomes, derived partly from \citet{bardzell_reading_2014}. The brief also specified an application context for the speculated near-future camera car: trash detection. We drew inspiration from an existing pilot project in Amsterdam. Garbage disposal may be a banal issue, but it is also multifaceted and has real stakes. We hired a filmmaker to collaborate with on video production. Funding for this part of the project came from AMS Institute, a public-private urban innovation center.\footnote{\url{https://www.ams-institute.org}} We first created a mood board to explore directions for the visual style. Ultimately, we opted for a collage-based approach because it is a flexible style that would allow us to depict complex actions without a lot of production overhead. It also struck a nice balance between accessibility and things feeling slightly off. We then wrote a script for the video. Here, we used contestability literature in general and the `Contestable AI by Design' framework \cite{alfrink_contestable_2022}, in particular, to determine what elements to include. We tried to include a variety of risks and related system improvements (rather than merely one of each) so that the audience would not quickly dismiss things for lack of verisimilitude. Having settled on a script, we then sketched out a storyboard. Our main challenge here was to balance the essential depiction of an intelligent system with potential risks, ways citizens would be able to contest, and the resulting system improvements. As we collaboratively refined the storyboard, our filmmaker developed style sketches that covered the most essential building blocks of the video.\footnote{Design brief, script, and storyboards are available as supplementary material.} Once we were satisfied with the storyboard and style sketches, we transitioned into video production. Production was structured around reviews of weekly renders. On one occasion, this review included partners from AMS Institute. Our next milestone was to get a rough cut of the video `feature complete' for assessment with experts.

For this assessment, we created an interview guide and a grading rubric. We based the rubric on the assessment criteria developed in the original design brief. All experts were colleagues at our university, selected for active involvement in the fields of design, AI and ethics. We talked to seven experts (five male, two female; two early-career researchers, three mid-career, and two senior). Interviews took place in early February 2022. Each expert was invited for a one-on-one video call of 30-45 minutes. After a brief introduction, we went over the rubric together. We then showed the concept video rough cut. Following this, the expert would give us the grades for the video. After this, we had an open-ended discussion to discuss potential further improvements. Audio of the conversations was recorded with informed consent, and (roughly) transcribed using an automated service. We then informally analyzed the transcripts to identify the main points of improvement. We first summarized the comments of each respondent point by point. We then created an overall summary, identifying seven points of improvement. We visualized the rubric score Likert scale data as a diverging stacked bar chart.\footnote{Interview guide, assessment form template, completed forms, tabulated assessment scores, and informal analysis report are available as supplementary material.}

Once we completed the expert assessment, we identified improvements using informal analysis of the automated interview transcripts. The first author then updated the storyboard to reflect the necessary changes. We discussed these with the filmmaker, and converged on what changes were necessary and feasible. The changes were then incorporated into a final cut, adding music and sound effects created by a sound studio and a credits screen.

\subsection{Civil servant interviews}

Interviews were conducted from early May through late September 2022. We used purposive and snowball sampling. We were specifically interested in acquiring the viewpoint of civil servants involved in using AI in public administration. We started with a hand-picked set of five respondents, whom we then asked for further people to interview. We prioritized additional respondents for their potential to provide diverse and contrasting viewpoints. We stopped collecting data when additional interviews failed to generate significantly new information. We spoke to 17 respondents in total. Details about their background are summarized in Table \ref{tab:demographics-respondents}. We invited respondents with a stock email. Upon expressing their willingness to participate, we provided respondents with an information sheet and consent form and set a date and time. All interviews were conducted online, using videoconferencing software. Duration was typically 30-45 minutes. Each interview started with an off-the-record introduction, after which we started audio recording with informed consent from respondents. We used an interview guide to help structure the conversation but were flexible about follow-up questions and the needs of respondents. After a few preliminary questions, we would show the video. After the video, we continued with several more questions and always ended with an opportunity for the respondents to ask questions or make additions for the record. We then ended the audio recording and asked for suggested further people to approach. After each interview, we immediately archived audio recordings and updated our records regarding whom we spoke to and when. We then sent the audio recordings to a transcription service, which would return a document for our review. We would review the transcript, make corrections based on a review of the audio recording where necessary, and remove all identifying data. The resulting corrected and pseudonymized transcript formed the basis for our analysis.\footnote{Interview guide is available as supplementary material.}

\begin{table*}[htb]
    \centering
    \begin{tabularx}{0.66666667\textwidth}{l X l}
        \toprule
        Item & Category & Number \\
        \midrule
        Gender & Female & 10 \\ 
               & Male & 7 \\
        Department & Digital Strategy and Information & 3 \\ 
                   & Legal Affairs & 2 \\
                   & Traffic, Public Space, and Parking & 2 \\
                   & Urban Innovation and R\&D & 10 \\
        Background & AI, arts \& culture, business, data science, information science, law, philosophy, political science, sociology & -- \\
        \bottomrule
    \end{tabularx}
    \caption{Summary of civil servant interview respondent demographics}
    \label{tab:demographics-respondents}
\end{table*}

\subsection{Analysis}

Our analysis of the data is shaped by critical realist \cite{frauenberger_critical_2016, green_algorithmic_2020} and contextualist \cite{jaeger_contextualism_1988, henwood_beyond_1994} commitments. We used reflexive thematic analysis \cite{braun_using_2006, cooper_thematic_2012, braun_thematic_2021} because it is a highly flexible method that readily adapts to a range of questions, data generation methods, and sample sizes. Because of the accessibility of its results, it is also well-suited to our participatory approach. The principal researcher took the lead in data analysis. Associate researchers contributed with partial coding and review of coding results. The procedure for turning ``raw'' data into analyzable form was: (1) reading and familiarization; (2) selective coding (developing a corpus of items of interest) across the entire dataset; (3) searching for themes; (4) reviewing and mapping themes; and (5) defining and naming themes. We conducted coding using Atlas.ti. We used a number of credibility strategies: member checking helped ensure our analysis reflects the views of our respondents; different researchers analyzed the data reducing the likelihood of a single researcher's positionality overly skewing the analysis; and reflexivity ensured that analysis attended to the viewpoints of the researchers as they relate to the phenomenon at hand.\footnote{Interview transcript summaries and code book are available as supplementary material.} In what follows, all direct quotes from respondents were translated from Dutch into English by the first author.

\section{Results}

\subsection{Concept video description}

\begin{figure*}[htb]
    \centering
    \includegraphics[width=1.0\textwidth]{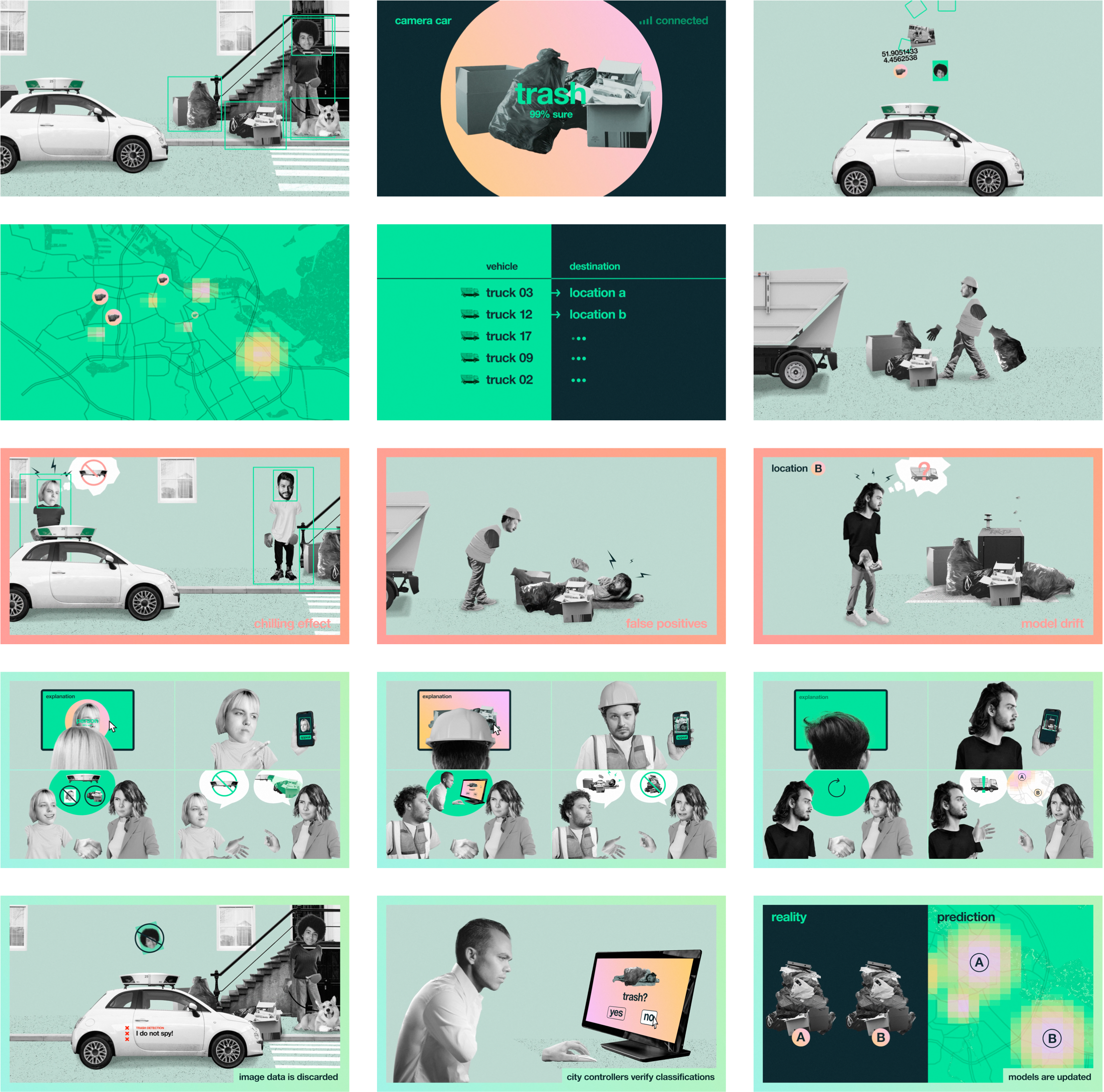}
    \caption{Stills from concept video}
    \Description[Stills from concept video.]{Stills from concept video. A camera car captures images of garbage and a person in the street. The car's computer vision system recognizes trash. The car sends data to the cloud. The system constructs a heat map of the city from camera car data. A prediction model determines where to send garbage collection vehicles. A sanitation worker collects garbage. People in the street dislike being spied on by a camera car and leave. A homeless person is shocked to be confronted by a sanitation worker. A person is confused about why garbage is piling up. Citizens consult explanations on system behavior. Citizens contact the city with questions. Citizens debate their concerns with city representatives. Citizens and the city decide how to respond to their concerns. The camera car discards data that is privacy sensitive. A person checks to ensure objects recognized by the camera car are trash. Data is updated to make the prediction model better fit reality.}
    \label{fig:concept-video-stills}
\end{figure*}

The concept video has a duration of 1 minute and 57 seconds. Several stills from the video can be seen in Figure \ref{fig:concept-video-stills}. It consists of four parts. The first part shows a camera car identifying garbage in the streets and sending the data off to an unseen place of processing. We then see the system building a heat map from identified garbage and a resulting prioritization of collection services. Then, we see garbage trucks driving off and a sanitation worker tossing the trash in a truck. The second part introduces three risks conceivably associated with the suggested system. The first risk is the so-called `chilling effect.' People feel spied on in public spaces and make less use of it. The second risk is the occurrence of `false positives,' when objects that are not garbage are identified as such, leading to wasteful or harmful confrontations with collection services. The third risk is `model drift.' Prediction models trained on historical data become out of step with reality on the ground. In this case, collection services are not dispatched to where they should be, leading to inexplicable piling up of garbage. The third part shows how citizens introduced in the risks section contest the system using a four-part loop. First, they use explanations to understand system behavior. Second, they use integrated channels for contacting the city about their concern. Third, they discuss their concern and point of view with a city representative. Fourth, the parties decide on how to act on the concern. The fourth and final part shows how the system is improved based on contestation decisions. The chilling effect is addressed by explicitly calling out the camera car's purpose on the vehicle itself, and personal data is discarded before transmission. False positives are guarded against by having a human controller review images the system believes are trash before action is taken. Finally, model drift is prevented by regularly updating models with new data. The video ends with a repeat of garbage trucks driving off and a sanitation worker collecting trash. A credits screen follows it.\footnote{The concept video is available as supplementary material.}

\subsection{Civil servant responses to concept video}

From our analysis of civil servant responses to the concept video, we constructed three themes covering 13 challenges. See Table \ref{tab:themes-overview} for a summary.

\begin{table*}[htb]
    \centering
    \begin{tabular}{l l l}
        \toprule
        Theme & \# & Challenge \\
        \midrule
        Enabling civic participation (\ref{sec:theme-1})
            & \ref{itm:citizen-capacities} & Citizen capacities \\
            & \ref{itm:communication-channels} & Communication channels \\
            & \ref{itm:feedback-to-dev} & Feedback to development \\
            & \ref{itm:reporting-inequality} & Reporting inequality \\
            & \ref{itm:participation-limitations} & Participation limitations \\
        Ensuring democratic embedding (\ref{sec:theme-2})
            & \ref{itm:democratic-control} & Democratic control \\
            & \ref{itm:external-oversight} & External oversight \\
            & \ref{itm:dispute-resolution} & Dispute resolution \\
        Building capacity for responsibility (\ref{sec:theme-3})
            & \ref{itm:organizational-limits} & Organizational limits \\
            & \ref{itm:accountability-infrastructure} & Accountability infrastructure \\
            & \ref{itm:civil-servant-capacities} & Civil servant capacities \\
            & \ref{itm:commissioning-structures} & Commissioning structures \\
            & \ref{itm:resource-constraints} & Resource constraints \\
        \bottomrule
    \end{tabular}
    \caption{Overview of themes and associated challenges}
    \label{tab:themes-overview}
\end{table*}

\subsubsection{Theme 1: Enabling civic participation}\label{sec:theme-1}

\begin{enumerate}[T{1}.1,listparindent=1em]
    \item \textit{Citizen capacities}\label{itm:citizen-capacities} (P1, P4, P5, P9, P10, P11, P12, P16, P17):
        Several respondents point out that contestability assumes sovereign, independent, autonomous, empowered, and articulate citizens. Citizens need sufficient awareness, knowledge, and understanding of systems to contest effectively.

        \begin{quote}
            \textit{``But everything actually starts with that information position as far as I am concerned.''} (P10)
        \end{quote}
        
        It can be hard for people to understand metrics used for evaluating model performance. For example, P17 describes how a model's intersection over union (IOU) score of 0.8 was talked about internally as an accuracy score of 80\%. Individuals also struggle to identify systemic shortcomings. Their view is limited to the impacts directly relating to themselves only. They may not even be aware that the decision that has impacted them personally was made in part by an algorithm. In addition, citizens can have false views of what systems do. For example, citizens and civic groups believed parking enforcement camera cars recorded visual likenesses of people in the streets, which was not the case. Citizens' ability to effectively contest further depends on how well they can navigate the city government's complicated internal organizational structure.

        Many respondents describe how citizens' willingness to engage depends on their view of city government. Those who feel the city does not solve their problems will be reluctant to participate. Citizens' inclination to scrutinize public algorithmic systems also depends on their general suspicion of technology. This suspicion appears to be at least somewhat generational. For example, younger people are more cautious about sharing their data. Suspicion is contextual, depending on what is at stake in a given situation. A lack of trust can also lead to citizens rejecting explanations and justifications offered by the city.

        \begin{quote}
            \textit{``I just think what a challenge it is to have a substantive conversation and how do you arrive at that substantive conversation.''} (P16)
        \end{quote}
        
    \item \textit{Communication channels}\label{itm:communication-channels} (P3, P4, P7, P8, P11, P12, P14, P16): 
        Many respondents recognize the importance of ensuring citizens can talk to a human representative of the city. Currently, citizens can contact the city about anything using a central phone number. Reports from citizens are subsequently routed internally to the proper channels.

        Ideally, the city should be able to route questions related to AI to civil servants who understand the relevant systems. Citizens are not able nor responsible for determining which issues pertain to algorithms and which do not. Triage should happen behind the scenes, as is currently the case with the central phone line. In other words, respondents would not favor a separate point-of-contact for `digital matters.'

        Executive departments are responsible for work processes, including those that use AI. They should therefore be the ones answering questions, \emph{including} those that relate to technology. But this is currently not always the case. Some respondents point out that development teams cannot be made responsible for answering citizens' questions. Despite this fact, these respondents describe how their development teams do receive emails from citizens and simply answer them.

        Beyond a central phone line, some respondents are considering other easily accessible, lower-threshold interaction modalities for expressing disagreement or concern (cf. Item \ref{itm:dispute-resolution}).
    
    \item \textit{Feedback to development}\label{itm:feedback-to-dev} (P1, P2, P3, P4, P5, P7, P10, P13, P14, P15, P17):
        Respondents feel it is important for development teams to seek feedback from citizens during development. 
        Indeed, for those systems developed internally, it is currently common practice to follow some iterative development methodology that includes testing pre-release software with citizen representatives.
        Most of the algorithmic systems discussed by respondents are still in this so-called pilot stage.
        Pilots are used to test new ideas for viability and explore the practical and ethical issues that might arise when a system is taken into regular everyday operation.
        
        \begin{quote}
            \textit{``But I also think testing is necessary for these kinds of things. So if you think it through completely, you will eventually see if you test whether it is feasible. Because now I have every time with such an iteration [...] you run into other things that make you think how is this possible?''} (P12)
        \end{quote}
        
        The city also conducts pilots to identify what is needed to justify the use of technology for a particular purpose.
        
        \begin{quote}
            \textit{``So we start a pilot in the situation where we already think: we have to take many measures to justify that. Because bottom line, we think it is responsible, but what do you think about this if we do it exactly this way? Do you agree, or is that [...] do you use different standards?''} (P7)
        \end{quote}

        Respondents involved with system development recognize that feedback from citizens can help eliminate blind spots and may lead to new requirements.

        Some respondents argue that all reports received by algorithmic system feedback channels should be open and public, or at least accessible to the municipal council so that democratic oversight is further enabled (cf. Item \ref{itm:democratic-control}). 
        
        On a practical level, to close the loop between citizens' reports and development, infrastructure is needed (cf. Item \ref{itm:accountability-infrastructure}). For example, the city's service management system, which integrates with the internal software development environment, is not yet open to direct reports from citizens but only from human controllers (cf. Item \ref{itm:accountability-infrastructure}). For those systems using machine learning models, there are no provisions yet for capturing feedback from citizens to retrain models (e.g., in a supervised learning approach).
    
    \item \textit{Reporting inequality}\label{itm:reporting-inequality} (P1, P4, P6, P12, P14, P15):
        Several respondents mention the issue of ``reporting inequality,'' where some citizens are more able and inclined to report issues to the city than others (cf. Item \ref{itm:citizen-capacities}). Some recent VUS efforts aim to counteract this reporting inequality; for example, the trash detection pilot our concept video took as a source of inspiration. Affluent neighborhoods are known to report on stray trash more than disadvantaged areas do and, as a result, are served better than is considered fair.

        Because of reporting inequality, respondents are weary of approaches that tie system changes directly to individual reports. For example, contestability may counteract the unequal distribution of vehicles due to system flaws, but it may just as well reintroduce the problem of reporting inequality. Contestability runs the risk of giving resourceful citizens even more outsize influence. Other respondents counter that making system changes in response to individual complaints may still be warranted if those changes benefit most citizens.

        Ultimately, many respondents feel it is up to developers and civil servants to interpret and weigh the signals they receive from citizens (cf. Item \ref{itm:feedback-to-dev}).

    \item \textit{Participation limitations}\label{itm:participation-limitations} (P1, P2, P4, P5, P6, P8, P10, P12, P14, P15, P16):
        Just as government should be aware of reporting inequality (cf. Item \ref{itm:reporting-inequality}), they should also ensure participation and contestation are representative. A real risk is that those with technical know-how and legal clout shape the debate around algorithmic systems. Respondents repeatedly point out that existing citizen participation efforts struggle with ensuring diversity, inclusion, and representation.

        \begin{quote}
            \textit{``For example, in [district], we also met someone who did many development projects with the neighborhood and who also agreed that, of course, the empowered people or the usual suspects often provide input and in [district] also low literacy and all sorts of other things make it much more difficult to [...] provide input if it is their neighborhood [...].''} (P2) 
        \end{quote}
        
        For the city, it is a struggle to find citizens willing and able to contribute to participation processes. Sometimes as a solution, the city compensates citizens for participating. Another way to improve inclusion is to go where citizens are rather than expect them to approach the city—for example, by staging events and exhibitions as part of local cultural festivals or community centers.

        Participation efforts assume direct representation. There is no mechanism by which individuals can represent interest groups. Citizens do not represent anyone but themselves and are not legally accountable for their decisions. Respondents point out that as one goes up the participation ladder \cite{arnstein_ladder_1969, cardullo_being_2017} more obligations should accompany more influence.

        Some respondents point out that government should take responsibility and depend less on individual citizens or hide behind participatory processes.
\end{enumerate}

\subsubsection{Theme 2: Ensuring democratic embedding}\label{sec:theme-2}

\begin{enumerate}[T{2}.1,listparindent=1em]
    \item \textit{Democratic control}\label{itm:democratic-control} (P1, P2, P3, P4, P5, P6, P7, P9, P10, P11, P12, P13, P14, P15, P16, P17): 
        Several respondents point out that the discretion to use AI for decision-making lies with the executive branch. For this reason, the very decision to do so, and the details of \emph{how} an algorithmic system will enact policy, should, in respondents' eyes, be a political one. Debate in the municipal council about such decisions would improve accountability.

        Respondents identify a tension inherent in public AI~projects: Policy-makers (alderpersons) are accountable to citizens and commission public AI projects, but they often lack the knowledge to debate matters with public representatives adequately. On the other hand, those who build the systems lack accountability to citizens. Accountability is even more lacking when developers do not sit within the municipal organization but are part of a company or non-profit from which the city commissions a system (cf. Item \ref{itm:commissioning-structures}).

        Respondents also point out that contestations originate with individual citizens or groups, but also with elected representatives. In other words, the municipal council \emph{does} monitor digital developments. The legislature can, for example, shape how the executive develops AI systems by introducing policy frameworks.

        P7 outlined three levels of legislation that embed municipal AI projects: (a) the national level, where the city must determine if there is indeed a legal basis for the project; (b) the level of local ordinances, which ideally are updated with the introduction of each new AI system so that public accountability and transparency are ensured; and finally (c) the project or application level, which focuses on the `how' of an AI system, and in the eyes of P7 is also the level where direct citizen participation makes sense and adds value (cf. Item \ref{itm:participation-limitations}).

        Feedback on AI systems may be about business rules and policy, which would require a revision \emph{before} a technical system can be adjusted.\footnote{This entanglement of software and policy is well-described by \citet{jackson_policy_2014}.} This then may lead to the executive adjusting the course on system development under its purview (cf. Item \ref{itm:feedback-to-dev}).

        There is also an absence of routine procedures for reviewing and updating existing AI systems in light of the new policy. Political preferences of elected city councils are encoded in business rules, which are translated into code. Once a new government is installed, policy gets updated, but related business rules and software are not, as a matter of course, but should be.

    \item \textit{External oversight}\label{itm:external-oversight} (P1, P2, P3, P4, P5, P6, P7, P8, P9, P10, P11, P12, P13, P14, P15, P16, P17):
        The city makes use of several forms of external review and oversight. Such reviews can be a requirement, or something the city seeks out because of, for example, citizens' lack of trust (cf. Item \ref{itm:citizen-capacities}).

        A frequently mentioned body is the local personal data commission (PDC). A PDC review is mandatory when a prospective algorithmic system processes personal data or when it is considered a high-risk application. The PDC focuses, among other things, on a system's legal basis, proportionality, and mitigation of identified risks. One respondent proposes that such human rights impact assessments be made open for debate.\footnote{Following widespread resistance against a 1971 national census, the Dutch government established a commission in 1976 to draft the first national privacy regulation. Because it collected and processed a significant amount of personal data itself, Amsterdam decided not to wait and created local regulations in 1980. Every municipal service and department was required to establish privacy regulations. The city established a special commission to review these guidelines and to decide if municipal bodies were allowed to exchange information, thereby creating the PDC (``Commissie Persoonsgegevens Amsterdam (CPA),'' {\url{https://assets.amsterdam.nl/publish/pages/902156/brochure_cpa_40_jarig_bestaan.pdf}}). The executive board expanded the tasks of the PDC in December 2021 ({\url{https://www.amsterdam.nl/bestuur-organisatie/college/nieuws/nieuws-19-januari-2022/}}). It now advises the board, upon request or on its own initiative, on issues ``regarding the processing of personal data, algorithms, data ethics, digital human rights and disclosure of personal data'' ({\url{https://assets.amsterdam.nl/publish/pages/902156/cpa_reglement.pdf}}). In the lead-up to this decision, in April 2021, a coalition of green, left and social liberal parties submitted an initiative proposal to the board that aimed to ``make the digital city more humane.'' It too argued for the expansion of the PDC's role ({\url{https://amsterdam.groenlinks.nl/nieuws/grip-op-technologie}}).}

        Other review and oversight bodies include the local and national audit offices, the municipal ombudsperson, and a so-called reporting point for chain errors. One shortcoming is that many of these are incident-driven. They cannot proactively investigate systems.

        Naturally, the civil servants, committee members, ombudspersons, and judges handling such cases must have a sufficient understanding of the technologies involved. External review bodies sometimes, at least in respondents' eyes, lack sufficient expertise. One example of such a case is recent negative advice delivered by a work participation council after a consultation on using AI by the work participation and income department to evaluate assistance benefit applications. At least one respondent involved in developing the system proposal felt that, despite considerable effort to explain the system design, the council did not fully grasp it.

        P7 considers judicial review by an administrative court of a decision that is at least in part informed by an algorithmic system, the ``finishing touch.'' When a client file includes the data that significantly impacts a model prediction, a judge's ruling on a municipal decision is implicitly also about the operation of the model.

        \begin{quote}
            \textit{``If [a decision] affects citizens in their legal position, for example, in the case of a fine [...] then yes, the administrative court can look into it. That is when it gets exciting. That is the finishing touch to what we have come up with.''} (P7) 
        \end{quote}
        
        This sentiment was echoed by P11 when they discussed how they could show in court what images the municipal parking monitoring camera car exactly captured, which received a favorable ruling from a judge.
    
    \item \textit{Dispute resolution}\label{itm:dispute-resolution} (P10, P11, P14, P15):
        Respondents feel that, for individual substantive grievances caused by algorithmic decision-making, existing complaint, objection, and appeal procedures should also work. These form an escalating ladder of procedures: complaints are evaluated by civil servants; objections go to an internal committee; if these fail, the case is handled by an ombudsperson; and finally, appeals procedures are handled by a judge.

        Respondents point out that existing procedures can be costly and limiting for citizens and not at all ``user-friendly.'' Existing procedures still rely heavily on communication by paper mail. Current procedures can be stressful because people are made to feel like an offender rather than being given the benefit of the doubt.

        \begin{quote}
            \textit{``And we criminalize the citizen very quickly if he does not want to—a difficult citizen, annoying. Yes, no, it is just that way, and no, sorry, bye. So there is little to no space, and if you have heard [a complaint] ten times from citizens, then maybe you should think about, we have ten complaining citizens. It is not one or two. There might be something wrong so let us look at that.''} (P13) 
        \end{quote}
        
        Respondents agree that more effort should be put into creating alternative dispute resolution mechanisms. These should help citizens stay out of costly and stressful legal proceedings. However, these ideas are mostly considered an `innovation topic,' which is to say, it is not part of daily operation. Such measures would require collaboration between those departments executing work processes and legal. At the moment, execution tends to consider dealing with disputes as not part of their remit. Legal does currently call citizens who have started an appeals procedure to make them feel heard, find alternative solutions, and offer them the opportunity to withdraw.

        Existing mechanisms do require more integration with technology. For example, case files should include all the relevant information about the data and algorithms used. Some services, such as parking monitoring, have already built custom web interfaces for appeals that integrate with algorithmic systems and offer citizens access to the data collected on their case. These would either expedite otherwise unwieldy legacy procedures or seek to keep citizens out of formal legal appeal procedures altogether.
\end{enumerate}

\pagebreak

\subsubsection{Theme 3: Building capacity for responsibility}\label{sec:theme-3}

\begin{enumerate}[T{3}.1,listparindent=1em] 
    \item \textit{Organizational limits}\label{itm:organizational-limits} (P4, P5, P7, P11, P15):
        Respondents point out that organizational fragmentation works against the city's capacity to respond to citizen reports. The problem is not necessarily that signals are not received by the city. Often the problem is that they are not adequately acted upon. Internal fragmentation also makes it hard for citizens to know who they should approach with questions (cf. Items \ref{itm:citizen-capacities}, \ref{itm:communication-channels}). For example, with parking, citizens are inclined to go to their district department, these need to pass on questions to parking enforcement, who in turn, if it concerns a street-level issue, must dispatch a community service officer.

        \begin{quote}
            \textit{``And I think that if you cut up the organization as it is now [...] then you might also have to work with other information in order to be able to deliver your service properly. So when we all had [more self-sufficient, autonomous] district councils in the past and were somewhat smaller, you could of course immediately say that this now has priority, we receive so many complaints or the alderman is working on it.''} (P11)
        \end{quote}
        
        Fragmentation and the bureaucratic nature of the city organization works against the adoption of `agile methods.'
        Although pilots are in many ways the thing that makes the innovation funnel of the city function, respondents also describe pilots as ``the easy part.'' The actual implementation in daily operations is a completely different matter. P3 describes this as the ``innovation gap.'' Transitioning a successful pilot into operation can easily take 3-5 years.

    \item \textit{Accountability infrastructure}\label{itm:accountability-infrastructure} (P2, P4, P5, P7, P11, P12, P13):
        Respondents discuss various systems that are put in place to improve accountability. The city is working to ensure requirements are traceable back to the person that set them, and developers record evidence to show they are met. Evidence would include email chains that record design decisions and system logging that shows specific measures are indeed enforced (such as deletion of data). Regarding models, respondents indicate the importance of validating them to demonstrate that they indeed do what they are said to do.

        Once past the pilot stage, monitoring and maintenance become essential considerations currently under-served. For this purpose, developers should correctly document systems in anticipation of handover to a maintenance organization. Systems must be ensured to operate within defined boundaries, both technical and ethical (impact on citizens), and the delivery of ``end-user value'' must also be demonstrated. Such monitoring and maintenance in practice require the system developers' continued involvement for some time.
     
        Another provision for accountability is the service management system integrated with the city's software development and operations environment (cf. Item \ref{itm:feedback-to-dev}). Several respondents point out that surveillance and enforcement are two separate organizational functions. For those AI systems related to surveillance and enforcement, a `human-in-the-loop' is currently already a legal requirement at the enforcement stage. Human controllers use the service management system to report system flaws, which may lead to changes and are fully traceable (cf. Item \ref{itm:feedback-to-dev}). Once in maintenance, with these systems in place, it should be possible for functional management to revise systems periodically, also in light of policy changes (cf. Item \ref{itm:democratic-control}).

        Several respondents argue that the city should also monitor individual complaints for issues that require a system change (cf. Item \ref{itm:feedback-to-dev}). 
    
    \item \textit{Civil servant capacities}\label{itm:civil-servant-capacities} (P1, P3, P4, P6, P7, P15): 
        Contestability puts demands on civil servants.
        
        \begin{quote}
            \textit{``[...] I think all contestability [shown in the video] assumes a very assertive citizen who is willing to contact a city representative who is willing to listen and has time for it and is committed to doing something about it.''} (P1) 
        \end{quote}
        
        Civil servants need knowledge and understanding of AI systems, including those employees that speak to citizens who contact the city with questions, e.g., through the central phone number. Politicians, city council members, and alderpersons also need this understanding to debate the implications of new systems adequately. At the level of policy execution, department heads and project leads are the ``first line of defense'' when things go wrong (P7). So they cannot rely on the expertise of development teams but must have sufficient understanding themselves. Finally, legal department staff must also understand algorithms. P15 mentions that a guideline is in the making that should aid in this matter.

        Beyond updating the knowledge and skills of existing roles, new roles are necessary. In some cases, agile-methods-style `product owners' act as the person that translates policy into technology. However, P7 feels the organization as a whole still lacks people who can translate legislation and regulations into system requirements. Zooming out further, respondents mention challenges with the current organizational structure and how responsibility and accountability require multidisciplinary teams that can work across technical and social issues (cf. Item \ref{itm:organizational-limits}).

    \item \textit{Commissioning structures}\label{itm:commissioning-structures} (P1, P3, P4, P11, P12, P13, P16, P17):
        The city can commission AI systems in roughly three ways, with different impacts on the level of control it has over design, development and operation: (a) by purchasing from a commercial supplier a service that may include an AI system; (b) by outsourcing policy execution to a third party, usually a non-profit entity who receives a subsidy from the city in return; or (c) by developing a system in-house.

        When purchasing, the city can exercise control mainly by imposing purchasing conditions, requiring a strong role as a commissioner. When out-placing policy execution, the city has less control but can impose conditions on the use of technology as part of a subsidy provision. When developing in-house, the city owns the system completely and is therefore in full control. In all cases, however, the city is `policy owner' and remains responsible for executing the law.

        These different collaboration structures also shape the possible dialogue between policy-makers and system developers at the start of a new project. When development happens in-house, an open conversation can happen. In the case of a tender, one party cannot be advantaged over others, so there is little room for hashing things out until an order is granted.

        Of course, collaboration with external developers can also have ``degrees of closeness'' (P4). More or less `agile' ways of working can be negotiated as part of a contract, which should allow for responding to new insights mid-course.

        Purchasing managers sometimes perceive what they are doing as the acquisition of a service that is distinct from buying technology solutions and can sometimes neglect to impose sufficient conditions on a service provider's use of technology.
        
        The duration of tenders is typically three years. On occasion, the city comes to new insights related to the responsible use of technologies a service provider employs (e.g., additional transparency requirements). However, it cannot make changes until after a new tender. Respondents point out that an additional feedback loop should lead to the revision of purchasing conditions. P17 describes a project in which parts of the development and operation are outsourced, and other components are done in-house. The decision on what to outsource mainly hinges on how often the city expects legislature changes that demand system updates.

    \item \textit{Resource constraints}\label{itm:resource-constraints} (P3, P4, P12, P16, P17): 
        Supporting contestability will require additional resource allocation. Respondents point out that the various linchpins of contestable systems suffer from limited time and money: (a) conducting sufficiently representative and meaningful participation procedures; (b) having knowledgeable personnel available to talk to citizens who have questions or complaints; (c) ensuring project leads have the time to enter information into an algorithm register; (d) performing the necessary additional development work to ensure systems' compliance with security and privacy requirements; and (e) ensuring proper evaluations are conducted on pilot projects.

        P12 compares the issue to the situation with freedom of information requests, where civil servants who are assigned to handle these are two years behind. Similarly, new legislation such as the European AI Act is likely to create more work for the city yet.

        For new projects, the city will also have to predict the volume of citizen requests so that adequate staffing can be put into place in advance. Having a face-to-face dialogue in all instances will, in many cases, be too labor-intensive (cf. Item \ref{itm:communication-channels}). A challenge with reports from citizens is how to prioritize them for action by city services, given limited time and resources (cf. Item \ref{itm:reporting-inequality}).
\end{enumerate}

\section{Discussion}

Our aim has been two-fold: (1) to explore characteristics of contestable public AI, and (2) to identify challenges facing the implementation of contestability in public AI. To this end, we created a speculative concept video of a contestable camera car and discussed it with civil servants employed by Amsterdam who work with AI.

\subsection{Summary of results}

\subsubsection{Concept video: Example of contestable public AI}

The speculative design concept argues for contestability from a risk mitigation and quality assurance perspective. First, it shows several hazards related to camera car use: chilling effect, false positives, and model drift. Then, it shows how citizens use contestability mechanisms to petition the city for system changes. These mechanisms are explanations, channels for appeal, an arena for adversarial debate, and an obligation to decide on a response. Finally, the video shows how the city improves the system in response to citizen contestations. The improvements include data minimization measures, human review, and a feedback loop back to model training. The example application of a camera car, the identified risks, and resulting improvements are all used as provocative examples, not as a prescribed solution. Together they show how, as \citet{alfrink_contestable_2022} propose, ``contestability leverages conflict for continuous system improvement.''

\subsubsection{Civil servant interviews: Contestability implementation challenges}

From civil servant responses to the concept video, we constructed three themes:
    \begin{enumerate}[{T}1]
        \item \textit{Enabling civic participation} (\ref{sec:theme-1}):
            Citizens need skills and knowledge to contest public AI on equal footing.
            Channels must be established for citizens to engage city representatives in a dialogue about public AI system outcomes.
            The feedback loop from citizens back to system development teams must be closed.
            The city must mitigate against `reporting inequality' and 
            the limitations of direct citizen participation in AI system development.
        \item \textit{Ensuring democratic embedding} (\ref{sec:theme-2}):
            Public AI systems embed in various levels of laws and regulations. An adequate response to contestation may require policy change before technology alterations.
            Oversight by city council members must be expanded to include scrutiny of AI use by the executive.
            Alternative non-legal dispute resolution approaches that integrate tightly with technical systems should be developed to complement existing complaint, objection, and appeal procedures.
        \item \textit{Building capacity for responsibility} (\ref{sec:theme-3}):
            City organizations' fragmented and bureaucratic nature fights against adequately responding to citizen signals.
            More mechanisms for accountability are needed, including logging system actions and monitoring model performance.
            Civil servants need more knowledge and understanding of AI to engage with citizens adequately. New roles that translate policy into technology must be created, and more multidisciplinary teams are needed.
            Contracts and agreements with external development parties must include responsible AI requirements and provisions for adjusting course mid-project.
            Contestability requires time and money investments across its various enabling components. 
    \end{enumerate}

\subsubsection{Diagram: Five contestability loops}

\begin{figure}[tb]
    \centering
    \includegraphics[width=0.9\columnwidth]{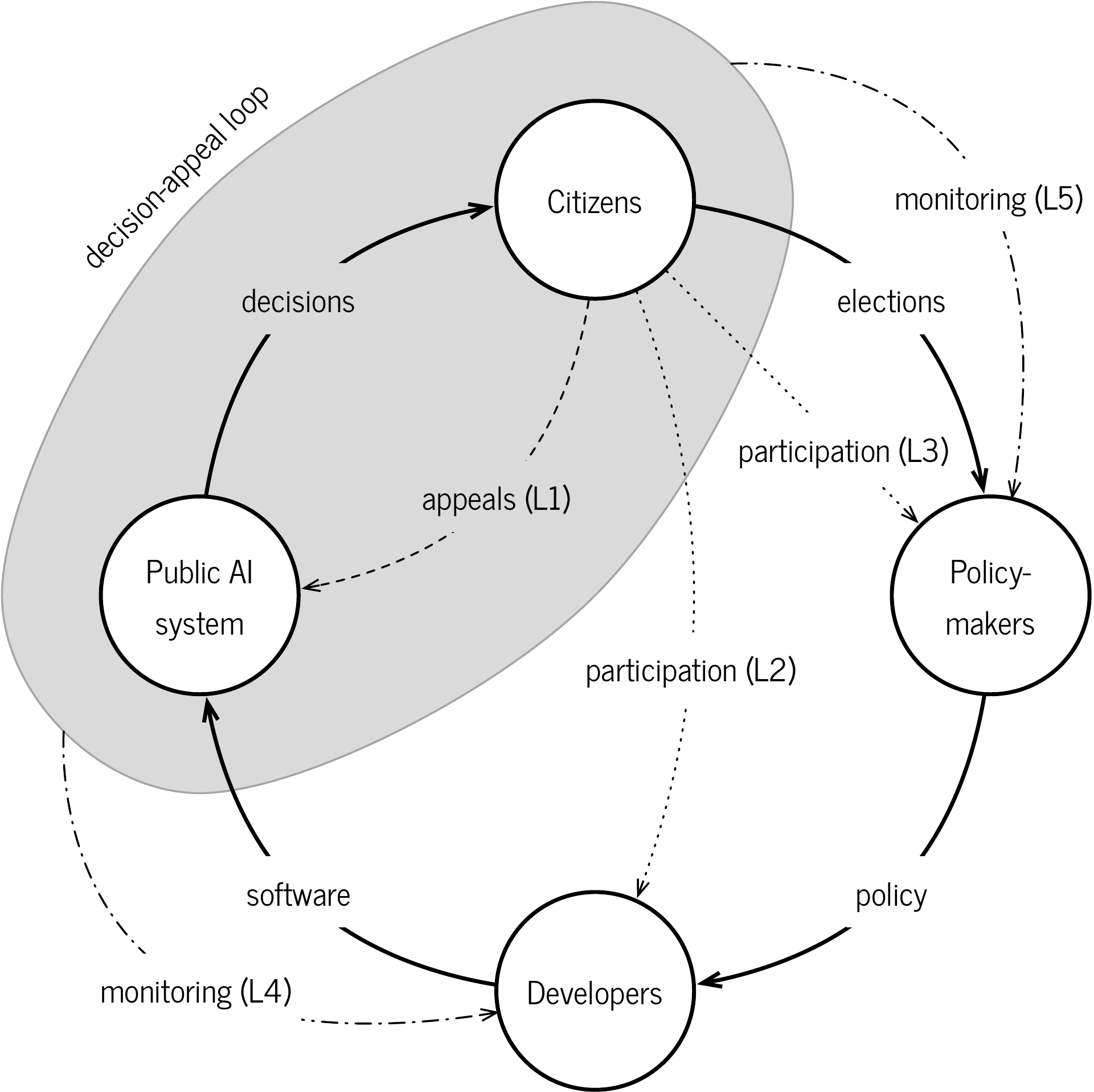}
    \caption{Diagram of our ``five loops model,'' showing the basic flow of policy through software into decisions (solid arrows), the direct way citizens can contest individual decisions (L1, dashed arrow), the direct ways in which citizens can contest systems development and policy making (L2-3, dotted arrows), and the second-order feedback loops leading from all decision-appeal interactions in the aggregate back to software development and policy-making (L4-5, dashed-dotted arrows).}
    \Description[Diagram showing the basic flow of policy through software into decisions, the direct ways in which citizens can contest these processes, and the second-order feedback loops leading from individual decision-appeal interactions back to software development and policy-making.]{Diagram showing the basic flow of policy through software into decisions, the direct ways citizens can contest these processes, and the second-order feedback loops leading from individual decision-appeal interactions back to software development and policy-making. Citizens elect policy-makers who develop policies for execution by a municipal organization that includes systems developers. These developers translate policy into software, data, and models. The resulting public AI systems make decisions that impact citizens. Citizens have three direct ways of contesting these processes: they can appeal an individual decision, participate in systems development, or participate in policy-making. The individual decision-appeal interactions are also monitored for systemic issues that require action from software development or policy-making.}
    \label{fig:concept-map}
\end{figure}

We can assemble five contestability loops from civil servants' accounts (Figure \ref{fig:concept-map}). This model's backbone is the primary loop where citizens elect a city council and (indirectly) its executive board (grouped as ``policy-makers''). Systems developers translate the resulting policy into algorithms, data, and models. (Other policy is translated into guidance to be executed by humans directly.) The resulting ``software,'' along with street-level bureaucrats and policy, form the public AI systems whose decisions impact citizens.

Our model highlights two aspects that are particular to the public sector context: (1) the indirect, representative forms of citizen control at the heart of the primary policy-software-decisions loop; and (2) the second-order loops that monitor for systemic flaws which require addressing on the level of systems development or policy-making upstream.

These five loops highlight specific intervention points in public AI systems. They indirectly indicate what forms of contestation could exist and between whom. To be fully contestable, we suggest that public AI systems implement all five loops. Better integration with the primary loop, and the implementation of second-order monitoring loops, deserve particular attention.

\subsection{Results' relation to existing literature}


\subsubsection{Contestable AI by design}

Following \citet{alfrink_contestable_2022}'s definition of contestable systems as ``open and responsive to human intervention,'' our respondents appear broadly sympathetic to this vision, particularly the idea that government should make more of an effort to be open and responsive to citizens.

We recognize many key contestability concepts in current city efforts as described by our respondents. For example, the possibility of human intervention \cite{hirsch_designing_2017} is mandatory in cases of enforcement, which can protect against model fallibility, at least to the extent errors can be detected by individual human controllers. Nevertheless, this human-in-the-loop is implemented more for legal compliance than quality control. Respondents talk about quality assurance and ways to achieve it, e.g., through audits and monitoring, but few practical examples appear to exist as of yet. The city recognizes the need to integrate institutional contestability provisions with technical systems (i.e., contestability as ``deep system property'' \cite{vaccaro_contestability_2019}). However, this integration is currently underdeveloped. Positive examples include the custom web interface for appealing parking enforcement decisions. Ex-ante contestability measures \cite{almada_human_2019} are present mainly in pilots in the form of civic participation in early-stage systems design. However, most participation happens on the project level and has no impact on policy decisions upstream from technology design. A dialectical relationship \cite{sarra_put_2020} is present on the far ends of what we could describe as the question-complaint-object-appeal spectrum; for example, the central phone line on one end and the review of algorithmic decisions by administrative courts on the other. The middle range seems to have less opportunity for exchanging arguments; again, these measures generally lack integration with technology. In any case, executing this ideal at scale will be costly. Finally, the city appears to approach accountability and legitimacy by ensuring the availability of explanations (e.g., in the form of an algorithm register). There appears to be less interest in, or awareness of, the need for justifications \cite{henin_beyond_2021} of decisions.

Most of the literature emphasizes contestability from below and outside but does not account for the representative democracy mechanisms in which public AI systems are embedded. In terms of our five loops model, city efforts emphasize individual appeals of decisions (L1) and direct participation in systems development (L2). Cities' policy execution departments are not, by their nature, adept at adjusting course based on external signals.

Furthermore, many cities still approach AI mostly from a pilot project perspective. Attitudes should shift to one of continuous learning and improvement. For example, Amsterdam conducts pilots with uncharacteristically high care. These pilots receive more scrutiny than systems in daily operation to allow for operation ``in the wild'' while staying within acceptable boundaries. The additional scrutiny throughout and the mandatory intensive evaluations upon completion serve to identify risks that may arise if systems were to transition into daily operation. This careful approach transforms pilots from the non-committal testing grounds common in the business world into something more akin to a social experiment guided by bioethical principles \cite{van_de_poel_ethical_2016}. While Amsterdam's pilots serve as good examples, successful pilots' transition into daily operation faces difficulties. This ``innovation gap'' (cf. Item \ref{itm:organizational-limits}) may be partially alleviated when designers stay involved after delivery. Public AI designers should consider themselves stewards, whose role is never finished \cite{dubberly_why_2022}.

Finally, it is not just AI and its development process that need `redesigning.' Cities' AI commissioning and governance structures must also be adjusted. Again referring to our five loops model, this would mean a focus on participation in policy-making (L3) and the second-order feedback loops from decision appeals to developers and policy-makers (L4-5).

\subsubsection{Public \& urban AI, and VUS}

Our example case of camera-car-based trash detection illustrates the need for the public and urban AI fields to converse more actively with each other. Public AI tends to focus on what goes on inside city organizations; urban AI tends to focus on what happens in the streets. Our results show how the concept of contestability connects the dots between several issues focused on in the literature so far. Namely, between explanations and justifications \cite{brown_toward_2019, dalpiaz_trustworthy_2021, de_fine_licht_artificial_2020}, street-level bureaucrat discretion \cite{flugge_perspectives_2021, saxena_conducting_2020, saxena_framework_2021}, and citizens' daily lived experience of urban space \cite{luusua_artificial_2020, luusua_urban_2020}.

Participation in public and urban AI literature is almost invariably of the direct kind \cite{saxena_conducting_2020, brown_toward_2019} as if we have given up on representative modes of democracy. There is potential in renewing existing forms of civic oversight and control. So, again, in our five loops model, a shift from focusing on individual appeals and direct participation in development (L1-2) to participation in policy-making (L3) and monitoring of appeals by policy-makers (L5).

We find it striking that the HCI design space appears to devote little or no attention to (camera-based) VUS. Camera cars appear to offer tremendous seductive appeal to administrators. More public camera car applications will likely find their way into the cities of the global north. They deserve more scrutiny from (critical) HCI scholars.

\subsubsection{Speculative design as a research method}

Turning to methodological aspects, we will make a few observations. As is often the case with contemporary speculative design, our concept is more a story than a product \cite{galloway_speculative_2018}. Indeed, we sought to spark the imagination of the audience \cite{sterling_cover_2009, dunne_speculative_2013}. One respondent recognized this: 

\begin{quote} 
    \textit{``And I think the lack of imagination that you have dealt with really well with your film is what keeps the conversation going even now, which is exactly the goal.''} (P9)
\end{quote}

The story we tell explores the implications of new technology \cite{dunne_speculative_2013}. It is a projection of potential future impacts of public AI that is (or is not) contestable \cite{dunne_speculative_2013}. Nevertheless, it would go too far to say we are `constructing a public' \cite{disalvo_design_2009}. We have not engaged in ``infrastructuring'' or the creation of ``agonistic spaces'' \cite{forlano_design_2014, galloway_speculative_2018, hillgren_prototyping_2011}. We \emph{did} design for one-on-one debate \cite{malpass_criticism_2015} and worked to ensure the video is sufficiently provocative and operates in the emotional register without tipping over into pure fancy or parody \cite{malpass_between_2013, malpass_criticism_2015}.

We used speculative design to open up rather than close down \cite{kozubaev_expanding_2020}. In this opening up, we went one step beyond merely critiquing current public AI practice and offered a speculative solution of contestability, framed in such a way that it invited commentary. Thus, asking questions rather than solving problems may not be the best way to distinguish speculative design from `affirmative design.' As \citet{malpass_criticism_2015} points out, rather than lacking function, critical design's function goes beyond traditional notions of utility, efficiency, and optimization and instead seeks to be relational, contextual, and dynamic.

On a more practical level, by building on the literature \cite{disalvo_design_2009, bardzell_what_2013, galloway_speculative_2018, bardzell_reading_2014}, we defined success criteria up front. Before bringing the result to our intended audience, we built an explicit evaluation step into our design process. This step used these same criteria to gain confidence that our artifact would have the effect we sought it to have on our audience. This approach can be an effective way for other design researchers to pair speculative design with empirical work.

\subsection{Transferability: Results' relation to city and citizens}

Amsterdam is not a large city in global terms, but populous and dense enough to struggle with ``big city issues'' common in popular discourse. Amsterdam was an early poster child of the ``smart cities'' phenomenon. It embraced the narrative of social progress through technological innovation with great enthusiasm. Only later did it become aware and responsive to concerns over the detrimental effects of technology. We expect that Amsterdam's public AI efforts, the purposes technology is put to, and the technologies employed are relatively common.

The city's government structure is typical of local representative democracies globally. Furthermore, the Netherlands's electoral system is known to be effective at ensuring representation. Many of the challenges we identify concerning integrating public AI in local democracy should be transferable to cities with similar regimes.

Amsterdam is quite mature in its policies regarding ``digital,'' including the responsible design, development, and operation of public AI. Less-advanced cities will likely struggle with more foundational issues before many of the challenges we have identified come into focus. For example, Amsterdam has made considerable progress concerning the transparency of its public AI system in the form of an algorithm register, providing explanations of global system behavior. The city has also made notable progress with developing in-house capacity for ML development, enabling it to have more control over public AI projects than cities dependent on private sector contractors.

Amsterdam's residents have a national reputation for being outspoken and skeptical of government. Indeed, city surveys show that a significant and stable share of the population is politically active. Nevertheless, a recent survey shows that few believe they have any real influence.\footnote{\url{https://onderzoek.amsterdam.nl/publicatie/amsterdamse-burgermonitor-2021}} Political engagement and self-efficacy are unequally divided across income and educational attainment groups, and these groups rarely encounter each other.

Our respondents tended to speak broadly about citizens and the city's challenges in ensuring their meaningful participation in public AI developments. However, in articulating strategies for addressing the challenges we have identified, it is vital to keep in mind this variation in political engagement and self-efficacy.

For example, improving citizens' information position so they can participate as equals may be relevant for politically active people but will do little to increase engagement. For that, we should rethink the form of participation itself. Likewise, improving the democratic embedding of public AI systems to increase their legitimacy is only effective if citizens believe they can influence the city government in the first place.

\subsection{Limitations}

Our study is limited by the fact that we only interacted with civil servants and the particular positions these respondents occupy in the municipal organization.

Over half of the civil servants interviewed have a position in the R\&D and innovation department of the city. Their direct involvement is mostly with pilot projects, less so with systems in daily operations. The themes and challenges we have constructed appear, for the most part, equally relevant across both classes of systems. It is conceivable, however, that civil servants employed in other parts of the city executive (e.g., social services) are more concerned with challenges we have not captured here.

Further work could expand on our study by including citizen, civil society, and business perspectives. This would surface the variety of interests stakeholder groups have with regards to contestability measures. Our respondents' statements are based on a first impression of the concept video. We expect more nuanced and richer responses if we give respondents more time to engage with the underlying ideas and apply them to their context. Finally, interviews do not allow for debate between respondents. Another approach would be to put people in dialogue with each other. This would identify how stakeholder group interests in contestability may align or conflict.

\subsection{Future work}

The public sector context brings with it particular challenges facing the implementation of contestability mechanisms, but also unique opportunities. For example, the existing institutional arrangements for contestation that are typical of representative democracies, on the one hand, demand specific forms of integration and, on the other hand, offer more robust forms of participation than are typically available in the private sector. For this reason, future work should include the translation of `generic' contestability design knowledge into context-specific forms. Considering the numerous examples of public AI systems with large-scale and far-reaching consequences already available to us, such work is not without urgency.

Most contestability research focuses on individual appeals (L1 in our five loops model) or participation in the early phases of AI systems development (L2, but limited to requirements definition). Future work should dig into the second-order loops we have identified (L4-5) and how citizens may contest decisions made in later phases of ML development (i.e., L2, but engaging with the `materiality' of ML \cite{benjamin_machine_2021, holmquist_intelligence_2017, dove_ux_2017}). The participatory policy-making loop (L3) is investigated in a more general form in, for example, political science. However, such work likely lacks clear connections to AI systems development implications downstream.

Finally, to contribute to public AI design practice, all of the above should be translated into actionable guidance for practitioners on the ground. Practical design knowledge is often best transmitted through evocative examples. Many more artifacts like our own concept video should be created and disseminated among practitioners. HCI design research has a prominent role in assessing such practical design knowledge for efficacy, usability, and desirability.

\section{Conclusion}

City governments make increasing use of AI in the delivery of public services. Contestability, making systems open and responsive to dispute, is a way to ensure AI respects human rights to autonomy and dignity. Contestable AI is a growing field, but the knowledge produced so far lacks guidance for the application in specific contexts. To this end, we sought to explore the characteristics of contestable \emph{public} AI and the challenges facing its implementation, by creating a speculative concept video of a contestable camera car, and conducting semi-structured interviews with civil servants who work with AI in a large northwestern European city. The concept video illustrates how contestability can leverage disagreement for continuous system improvement. The themes we constructed from the interviews show that public AI contestability efforts must contend with limits of direct participation, ensure systems' democratic embedding, and seek to improve organizational capacities.

`Traditional' policy execution is subject to scrutiny from elected representatives, checks from the judiciary and other external oversight bodies, and direct civic participation. The shift to AI-enacted public policy has undermined and weakened these various forms of democratic control. Our findings suggest that contestability in the context of public AI does not mean merely allowing citizens to have more influence over systems' algorithms, models, and datasets. Contestable \emph{public} AI demands interventions in how executive power uses technology to enact policy.

\begin{acks}
    We thank Roy Bendor, for advising us on our method and approach; Thijs Turèl (Responsible Sensing Lab and AMS Institute) for supporting the production of the concept video; Simon Scheiber (Trim Tab Pictures) for creating the concept video; the interviewed experts for their productive criticism that led to many improvements to the concept video; the interviewed civil servants for taking the time to talk to us and providing valuable insights into current practice; and the reviewers for their constructive comments. This research was supported by a grant from the Dutch National Research Council NWO (grant no. CISC.CC.018). 
\end{acks}

\bibliographystyle{ACM-Reference-Format}
\bibliography{bibliography}


\end{document}